# Ferroelectric driven exciton and trion modulation in monolayer MoSe$_2$ and WSe$_2$

*Bo Wen[1,2]\*, Yi Zhu[2,3]\*, Didit Yudistira[4], Andreas Boes[4], Linglong Zhang[2], Tanju Yidirim[5], Boqing Liu[2], Han Yan[2], Xueqian Sun[2], Yu Zhou[2], Yunzhou Xue[5], Yupeng Zhang[1], Lan Fu[3], Arnan Mitchell[4]★, Han Zhang[1]★, Yuerui Lu[2]★*

[1]Collaborative Innovation Center for Optoelectronic Science and Technology, International Collaborative Laboratory of 2D Materials for Optoelectronic Science and Technology of Ministry of Education and Guangdong Province, College of Optoelectronic Engineering, Shenzhen University, Shenzhen 518060, China.

[2]Research School of Engineering, College of Engineering and Computer Science, the Australian National University, Canberra, Australian Capital Territory 2601, Australia.

[3]Department of Electronic Materials Engineering, Research School of Physics and Engineering, The Australian National University, Canberra 2601, ACT, Australia.

[4]School of Engineering, RMIT University, Melbourne, Victoria 3001, Australia.

[5]College of Chemistry and Environmental Engineering, Shenzhen University, Shenzhen 518060, Guangdong, China.

\*These authors contributed equally to this work.

★E-mail: Yuerui Lu (yuerui.lu@anu.edu.au), Han Zhang (hzhang@szu.edu.cn) and Arnan Mitchell (arnan.mitchell@rmit.edu.au)

**Abstract**

In this work, we show how domain engineered lithium niobate can be used to selectively dope monolayer MoSe$_2$ and WSe$_2$ and demonstrate that these ferroelectric domains can significantly enhance or inhibit photoluminescence (PL) with the most dramatic modulation occurring at the heterojunction interface between two domains. A micro-PL and Raman system is used to obtain spatially resolved images of the differently doped transition metal dichalcogenides (TMDs). The domain inverted lithium niobate causes changes in the TMDs due to electrostatic doping as a result of the remnant polarization from the substrate. Moreover, the differently doped TMDs (n-type MoSe$_2$ and p-type WSe$_2$) exhibit opposite PL modulation. Distinct oppositely charged domains were obtained with a 9-fold PL enhancement for the same single MoSe$_2$ sheet when adhered to the positive (P$^+$) and negative (P$^-$) domains. This sharp PL modulation on the ferroelectric domain results from different free electron or hole concentrations in the materials conduction band or valence band. Moreover, excitons dissociate rapidly at the interface between the P$^+$ and P$^-$ domains due to the built-in electric field. We are able to adjust the charge on the P$^+$ and P$^-$ domains using temperature *via* the pyroelectric effect and observe rapid PL quenching over a narrow temperature range illustrating the observed PL modulation is electronic in nature. This observation creates an opportunity to harness the direct bandgap TMD 2D materials as an active optical component for the lithium niobate platform using domain engineering of the lithium niobate substrate to create optically active heterostructures that could be used for photodetectors or even electrically driven optical sources on-chip.

Direct bandgap monolayer transition metal dichalcogenides (TMDs)[1-4] have attracted tremendous interest due to their optically controlled valley polarization and coherence,[5-6] giant spin-valley coupling and tightly bound excitonic states.[7-10] Owing to their atomically thin structure, monolayer TMDs act as semiconductors which can undergo a complete transition from indirect to a direct bandgap, with energy gaps located at the Brillouin zone producing a strong exciton pumping efficiency.[2] In particular, monolayer TMDs provide a platform for investigating the dynamics of excitons in reduced dimensions at room temperature and fundamental many-body interactions.[11-17] The ability to access exciton behaviors at room temperature with this platform creates opportunities for high quality optical detection and ultimately optical sources.[18] So far, many investigations for TMD p-n diodes (positive-negative) consist of one monolayer TMD material and two splitting gate metals.[19-21] This method sets a gate voltage through the metallic contacts to dope the TMDs achieving strong charge-density tuning and therefore creates a lateral p-n junction. However, the metal gates can result in an inhomogeneous charge distribution and unavoidable quenching of light emission at the TMD surface.

Owing to ultra-thin nature of TMDs TMD materials are sensitive to the surrounding ionic environment, which opens the possibility of ferroelectric gating control of monolayer TMDs due to quantum confinement, high carrier mobility and a tunable bandgap. Ferroelectric (FE) materials possess a spontaneous electrical polarization resulting in a strong built in electric field along polar axis and very strong surface charges with positive charges on one surface and negative charges on the other. The spontaneous electrical polarization can be inverted through application of a strong electric field and this can be done locally to create a spatial domain pattern of positive or negative polar surfaces. This domain pattern can offer a strategy for lateral modulation in TMDs to create electro-statically driven p-n homojunction, which offers an alternative to complex split-gate electrodes.[22] Importantly, if the ferroelectric substrate is optically transparent, then the quenching of light emission often encountered with split-gate electrodes can be overcome.

Lithium niobate ($LiNbO_3$, LN) is a widely known ferroelectric material for optical waveguides, optical modulators, piezoelectric sensors and is an industry standard platform for wavelength conversion, high speed communications, microwave photonics and emerging platform for quantum optics.[23-27] Its utility in photonics is based on its excellent optical transparency from visible to mid infrared wavelengths and its strong ferroelectric nature leads to relatively efficient and ultra-high speed electro-optic and nonlinear optic properties. Due to the ability to integrate so many functionalities LN has recently been proposed as a next generation optical integration platform.[26] However, since LN is not a semiconductor, it cannot inherently provide high speed photodetectors or laser sources. This has led to many researchers exploring hybrid integration to obtain these functionalities.[25]

Hybridization of TMDs with LiNbO$_3$ thus presents a tantalizing opportunity to introduce electronically driven active optical functionality to the LN platform – essentially completing the toolbox of functional components that can be integrated on a single chip. By domain engineering patterns into a LN substrate it could be possible to modulate the carrier density in a monolayer TMD and create optically active p-n junctions – with geometry defined by the substrate poling pattern - independent of the alignment of the TMD film.[28-29] To pursue this vision, there is still a great deal of fundamental research to be done. The atomistic details and environment (temperature, pressure *etc*.) governing the interaction and the resulting charge density changes inside TMDs still remains elusive. In addition, poly-domain ferroelectrics and their interactions with TMDs have received limited theoretical and experimental attention. The impact of ferroelectric polarization in monolayer TMDs is of high interest and may open up potential possibilities in nanotechnology devices.

Herein, we explore the interaction of domain engineered LN with monolayer MoSe$_2$ and WSe$_2$ and observe dramatic modulation of the PL intensity and spectra. The LN substrates are domain engineered with an arbitrarily chosen periodic hexagonal patterns to illustrate the ability to engineer domains in arbitrary 2D configurations.[30] Using a confocal micro-PL system, TMDs (n-type MoSe$_2$ and p-type WSe$_2$) with different initial doping show opposite PL modulation mappings on LN substrates, demonstrating distinct behaviors on the two domain orientations. Moreover, an intriguing p-n homojunction platform in both monolayer MoSe$_2$ and WSe$_2$ integrated on the LN substrate was observed with distinct PL behavior observed at the interface between two domains. The LN substrate provides a non-volatile control of TMD doping and produces a p-n homojunction band structure. LN is pyroelectric with the surface charge being strongly temperature dependent. We show that small changes in temperature can result in dramatic changes in PL emission, illustrating the electronic nature of the PL phenomenon and suggesting that similar modulation could be achieved electronically. Experimental observations of the PL intensity modulation in TMD/LN match the intuitive predictions of compensation for ferroelectric polarization. This technology is also suitable for other types of 2D materials.[31-37] We believe that these results will open an exciting new avenue for LN optoelectronic components and in particular may enable electrically driven optical sources to be integrated on the LN photonic chips.

**Results and discussion**

A schematic plot of a TMD/LN homojunction is shown in Figure 1a and the fabrication process is summarized in the Methods section. The optical microscope images of the MoSe$_2$ and WSe$_2$ on domain engineered LN are shown in Figures S1b and c, respectively. As indicated in the scheme plot of Figure 1b, the LN substrate domains can be used to monitor the doping level and density in a TMD. On the up-domain state (P$^+$), the substrate is rich in negative interface charges and likely to enhance the intrinsic n-type character of MoSe$_2$. On the down-domain state (P$^-$) of the LiNbO$_3$, the interaction between the MoSe$_2$ and the substrate has the opposite effect.

The domain engineered LN substrate has different polarization states on the surface however, the optical properties are not modified making it difficult to locate the hexagonal structures from microscope directly. To overcome this challenge, a phase-shifting interferometer (Vecco NT9100) has been used to characterize the number of layers of the TMD samples and for the location of the hexagonal domains in the LN. The different surface charges on the two LN domains modify the index of the TMD films differently making it possible to observe the locations of the domains as small changes in phase shift through the TMD films. The spatial distribution of the hexagonal structures was precisely identified by phase-shifting interferometry (PSI) (Figures 1c and d). On the basis of the intuitive picture of the up and down polarized domains, we expect that these poly-domain states could create periodic spatial carrier charge modulation across the interface of the TMD/LN. The PSI was also used to identify the number of layers of the TMDs[38]. Owing to the multiple interfacial light reflections, the optical path length (OPL) of the light reflected from the TMDs determines the layer number precisely. As shown from the green box in Figure 1c, the uniform OPL yield indicates a good quality of TMD sample after it had been transferred to the substrate.

After the PSI measurement, LN chips were placed into a microscope-compatible chamber (Linkam THMS 600) for photoluminescence (PL) measurements to spatially probe the carrier density in the TMDs by optical method (Figure 2). The spectroscopy measurements were taken using a confocal microscopic setup with an excitation laser wavelength of 532 nm. Typical emission spectra and PL mapping of the monolayer TMD samples were recorded under an optical pump power at about 18 μW and the results are presented in Figure 2. For comparison, the results of the PL mapping from two different TMD samples are presented in Figures 2a and 2c, respectively. Normally, monolayer TMDs have large PL emission performance due to large quantum yield of radiative exciton recombination. The large binding energy of these photo-excited electrons and holes can form excitons and/or trions at room temperature. A trion is a charged exciton composed of two electrons and one hole or opposite composition, analogous to $H^-$ (or $H^{2+}$).

Inside the $P^-$ domain area, the PL emission of the monolayer $MoSe_2$ was enhanced 9 times compared to the $P^+$ domain as shown in the spatial modulation image in Figure 2a and confirmed by the PL spectra as shown in Figure 2b. The PL of the monolayer $WSe_2$ shows the opposite PL emission performance for both polarized domains compared to the $MoSe_2$ (Figure 2c). The PL modulation of the TMDs from the $P^-$ to $P^+$ domain shows a sharp transition at the ferroelectric domain boundary. On the basis of the PL mapping, the polarization orientation of the 2D domain patterns create regions of different charge carrier density in the TMD samples; the spatially modulated PL mappings revealed increased and decreased electron concentrations in the TMD samples. Due to the different doping types of the semiconductors, monolayer $WSe_2$ and $MoSe_2$ samples presented opposite PL intensity modulation after their integration with domain engineered LN. As schematically shown in Figure 2e, the doping level is changed to the opposite direction on different polarized domains. The enhanced PL emission of $MoSe_2$ in the $P^-$ area indicates the low doping level of $MoSe_2$, which is attributed to decreased n-type carrier concentrations in the $MoSe_2$.[39-41] Photoluminescence mapping revealed that periodic regimes of quenched/enhanced intensity are directly correlated with inverted ferroelectric polarization. This domain engineered ferroelectric substrate can provide simple emission modulation in TMD materials without the need for complex electrodes or multiple layers of semiconducting material.

In order to investigate surface doping that originates from the ferroelectric polarization effects due to monolayer PL emission, monolayer $MoSe_2$ and $WSe_2$ metallic oxide semiconductor field effect transistor (MOS) devices were fabricated to obtain the electrically gated PL spectrum (Figure S2 and S3). Monolayer $MoSe_2$ and $WSe_2$ were exfoliated onto $SiO_2/Si$ substrates from the same matrix crystal to keep the initial doping level the same. The initial doping level, surface doping level and photo doping level in monolayer TMDs were extracted by comparing PL spectra from the mapping data with the gate-dependent PL spectra from the MOS devices. Here, a monolayer $WSe_2$ was used to quantitively study charge transfer mechanisms. The gate dependent PL spectrum of monolayer $WSe_2$ is shown in Figure S2. By fitting all of the PL spectra under different gate voltages, the exciton and trion peak energies can be extracted. Then the Fermi level can be calculated by

$$E_A - E_T = E_F + E_{binding} \quad , \quad (1)$$

where $E_A$ is the exciton peak energy, $E_T$ is the trion peak energy, $E_F$ is the Fermi energy level and $E_{binding}$ is the binding energy of trion.

Fermi level of monolayer $WSe_2$ was extracted through Equation (1). By plotting energy difference betwee the exciton and trion peak as a function of gate voltage linear trend between energy difference and Fermi level can be observed. For monolayer TMDs, the density of state in the conduction band and valence band is linearly distributed along the energy level.[42] The extracted binding energy of the monolayer WSe$_2$ was 34 meV, consistent with previously reported results.[34, 43] To extract the doping level of the monolayer TMDs from the experimentally obtained Fermi level, the following two equations were used

$$E_F = \frac{\hbar \pi n}{2m_h e^2} , \qquad (2)$$

$$ne = CV_g , \qquad (3)$$

where $h$ is Planck's constant, $\pi$ is the pi constant, $n$ is the electron density, $e$ is the fundamental unit of charge, $m_h$ is the effective mass of a hole, $C$ is the back-gate capacitance and $V_g$ is the gate voltage.

Equation (2) and (3) were used to extract the doping level $n$ of the monolayer TMDs from the Fermi level and to convert the gate voltage into the induced hole doping level, respectively, where $C$ is the back-gate capacitance $C = 1.2 \times 10^{-8}\ F cm^{-2}$. As shown in Figure S2b and S2d, monolayer WSe$_2$ will approach a zero Fermi level when a -20V gate voltage is applied and this means monolayer WSe$_2$ is slightly p-type doped with an initial doping level of $1.5 \times 10^{12}\ cm^{-2}$ (when no gate voltage is applied). In Figure S4a and S4b, the value of $E_A - E_T$ for PL enhancement and quenching have been extracted as 31.47 meV and 41.41 meV, respectively. According to the linear relationship between Fermi level and $E_A - E_T$, the Fermi level and hole doping level for the P$^+$ domain are 0.92 meV and $2.54 \times 10^{11}\ cm^{-2}$, respectively, while for the P$^-$ domain they are 15.39 meV and $4.24 \times 10^{12}\ cm^{-2}$, respectively. Considering initial doping of monolayer WSe$_2$, the P$^+$ domain has induced $1.25 \times 10^{12}\ cm^{-2}$ electrons and the P⁻ domain has induced $2.74 \times 10^{12}\ cm^{-2}$ holes. Overall, the P⁻ domain has a lower hole doping level than the P⁺ domain meaning the P⁻ domain will have a higher density of vacant state in the valence band for PL emission (Figure 2e). The more neutral the monolayer TMD, the stronger PL emission.[40-41, 44]

The surface doping can effectively modify the doping level of two-dimensional materials[28]. Although, the optical behavior within the P⁻ and P⁺ domains have been subject to detailed investigations in the above discussion; the adjacent vicinity at the contact of the P⁻ and P⁺ domains (of the domain engineered LN and TMD), however, still has not been fully explored yet. In Figure 3, power dependent PL measurements were employed to unveil the behavior of the surface junction area. Figure 3a and 3b present confocal PL mapping under 1.8 and 178 μW, respectively. The PL intensity evolution increased with the laser power excitation as expected. As the excitation power increased, the PL intensity of junction area P2 dramatically increased compared with the P⁻ domain (P1) and P⁺ domain (P3), which the line curve is also presented in Figure 3d. In the P2 domain, the surface doping level is near zero because negative and positive charges are neutralized forming a depletion region and an internal built-in electrical field exists in this domain (Figure 3c). When the laser excitation power is very small, only a small number of photons can be injected into the material. Assuming a quantum efficiency of 1, only a small number of excitons are generated in this domain. They are quickly dissociated because of the internal built-in electric field.[45] Therefore, the P2 domain has lower PL emission compared with the P⁻ and P⁺ domains when the excitation power is 1.8 μW. However, when a larger excitation power is used, more excitons are formed and also dissociated. Electrons and holes accumulate at the edge of this depletion region. These accumulated charges create an electron/hole concentration gradient toward the depletion region, which is opposite to the built-in electrical field. These concentration gradients stop the dissociation process of excitons within the P2 area and eventually they reach a dynamic balance. When the laser power was increased to 178 μW, more excitons are generated in all P1, P2 and P3 areas. However, the material doping level for P2 is much smaller than P1 and P3 due to the depletion region. As shown in the scheme plot of Figure S5, due to the lower interface charge density of P2, the exciton in this junction area would increase more than other domains with increasing laser power (Figure S6a). This lower doping level and more unoccupied electron density in the conduction band can contribute to more PL emissions excited by larger laser power. Due to the lower doping level, PL performance in P2 area shows higher exciton emission efficiency, as shown in Figure S6b. With abundant electrons on the surface of P3 state, the p-type $WSe_2$ shows lower trion emission with increasing laser power as shown in in Figure S6b. With the balance of the built-in electric field, the intensity ratio between exciton and trion shows continuous reduction with increasing laser power, as shown in Figure S6c. Hence, as shown in Figure 3d, the PL emission in P2 is much stronger than P1 and P3.

Monolayer TMD samples on LN shows sizable interaction across the interface which affects the doping level and carrier density due to the remnant surface polarization of the FE substrate. Furthermore, due to the asymmetric polarization structure, the LN has an inimitable pyroelectric effect with changing temperature.[46] The surface interaction of ferroelectric LN and monolayer TMD materials at low temperature is largely unexplored. To study the thermal stability and spatial charge modulation properties of the TMD/LN, temperature dependent PL measurements with an excitation laser power of 18 μW have been conducted. The surface charge density may reduce due to absorption or reconstruction within the TMD/LN system with decreasing temperature.

As shown in Figure 4a, PL mapping demonstrates the emission characteristics at near room temperature. By reducing the temperature only slightly from 301 $K$ to 283 $K$, the PL intensity within the P$^-$ domain is rapidly quenched by nearly 75% (Figure S7). The PL intensity continues decreasing with decreased temperature down to 273 $K$ (Figure 4b), then the PL intensity keeps almost stable until liquid nitrogen temperature (83 $K$) (Figure 4b). The temperature dependent PL of MoSe$_2$ on LiNbO$_3$ are quite different from the PL performances of monolayer MoSe$_2$ on SiO$_2$, which increases with decreasing temperature (Figure S8a). Temperature changes can significantly affect the spatial carrier concentration of the TMD samples through the impact of the pyroelectric effect on spontaneously polarized surface charges.

The remnant polarization of the LiNbO$_3$ increases quickly with decreasing temperature from room temperature (RT) to 247$K$.[47] As illustrated in Figure S10, the interface charges would also increase quickly. The electron density in the MoSe$_2$ in the P$^-$ domain rapidly reduces and transfers to the substrate around RT to 247$K$. The doping level of MoSe$_2$ on different domains would form a device similar to a p-n junction. With decreasing temperature, interface charges in the P$^-$ and P$^+$ domains (hole and electron) would both increase, which is likely to apply a voltage to the overlayer MoSe$_2$ (Figure 4f). Combined with the MoSe$_2$ in the P$^+$ domain, the majority carriers (as holes in MoSe$_2$ on P$^-$ domain) would transfer to the nearby domains, which balances the charge density of different domains and the PL would also be balanced between different domains. As shown in Figure 4c, the PL intensity on the P$^-$ domain is quickly reduced and becomes similar to the PL intensity on the P$^+$ domain. The PL intensity of WSe$_2$ on the P$^+$ domain also decreased with decreasing temperature down to 83 $K$ and equals the PL emission on the P$^-$ domain (Figure 4d and e). When the temperature decreased from 247 $K$ to 83 $K$, the polarization of the substrate also decreased (Figure S9). So, the PL emission is slightly recovered and shown a hump around 220 $K$ as shown in Figure 4c and e. With continuous cooling, the distance between the overlayer MoSe$_2$ and the substrate is decreasing,[48-49] which enhanced the interaction between TMD samples and the FE substrate. Therefore, the overlayer transport properties between different domains are enhanced and quenched the PL emission on the P$^-$ domain. The recovery (the hump around 220 $K$) attributed to the coupling effect of enhanced interface charge transfer and decreased polarization of the LN substrate, which delayed the inflection point from 247 $K$ to 220 $K$. The PL emission intensity is rapidly quenched in the P$^-$ region due to the pyroelectric effect of LN (Figure 4f). The decreasing speed was indicated by the quench factor as shown in Figure S10. This observation clearly illustrates the electronic nature of the observed PL behavior. Due to the electrostatic doping as a result of the remnant polarization from the substrate, the band structure of monolayer MoSe$_2$ has no or little change. So the peak energy is almost the same on P$^-$ and P$^+$ domain (Figure S11b). Due to the electrostatic doping nature across the interface, little difference of peak energy and full width at half maxima (FWHM) of monolayer MoSe$_2$ on LN and SiO$_2$ is indicated, as shown in Figure S11d and S11e. The temperature dependent PL could be harnessed directly as a highly sensitive temperature sensor, however we believe that the insight into the optoelectronic behavior of the TMD material on the domain engineered LN presents a far more significant opportunity for electronically powered and controlled optical sources on the LN integrated photonic platform.

**Conclusion**

In conclusion, we have shown that monolayer transition metal dichalcogenides (TMD) MoSe$_2$ and WSe$_2$ can be interfaced onto the surface of ferroelectric lithium niobate (LN) and that the domain orientation of the LN has a strong effect of the TMD optoelectronic properties. Domain engineering of the LN substrate results in an in-plane heterostructure within the TMD which has been demonstrated by our measured spatial PL mapping. Due to the underlying domain engineered ferroelectric, PL emissions are modulated approximately 9 times by different surface doping between the P$^-$ and P$^+$ domains. Monolayer WSe$_2$ on LN also shows enhanced PL emission at the junction between domains with increasing laser power, this is due to overcoming the micro internal electric field between the P$^-$ and P$^+$ domains, which stops dissociation of excitons. With decreasing temperature, the PL emission of monolayer MoSe$_2$ was quickly quenched in the P$^-$ domain in stark contrast to behavior observed on non-ferroelectric substrates. This performance shows the fast increase of positive charge in the P⁻ domain due to the pyroelectric effect and the restructure of the surface doping between $MoSe_2$ and the LN substrate. Overall, our research shows that monolayer TMD sheets can be engineered to create a p-n homojunction by engineering the underlying ferroelectric domains with sensitive response to laser power and ambient temperature. This new approach could lay the foundations for creating active electrically driven and controlled optoelectronic components on LN integrated photonic platforms with the ultimate goal of electrically driven optical sources within LN photonic chips.

**Experimental Section**

**Device fabrication and characterization.** Monolayer TMD samples were produced by mechanical exfoliation from bulk crystal. Then, monolayer TMD samples were dry transferred onto a 500-μm-thick Z-cut LN substrate with periodical honeycomb lattice and hexagonal shape of the domain-inverted inclusions. Hexagonal inclusions are fabricate in trigonal (3m) LN crystal by electric field poling. The inclusion boundaries are along the symmetry planes in the *Z*-cut. The side length of the hexagon is about 6.16 μm. The PPLN substrate was cleaned with $O_2$ plasma as pretreatment, using a plasma cleaner. For the MOS structure, we also used mechanical exfoliation to dry transfer the TMD flake onto a $SiO_2$/Si substrate (275nm thermal oxide on n⁺-doped silicon), with half on a pre-patterned gold electrode and half on the substrate. The gold electrodes were patterned by conventional photolithography, metal deposition and lift-off processes.

**Optical characterization.** All the OPL characterizations were obtained using a phase-shifting interferometer (Vecco NT9100). PL measurements were conducted using a Horiba LabRAM system equipped with a confocal microscope, a charge-coupled device (CCD) Si detector, and a 532 nm diode-pumped solid-state (DPSS) laser as the excitation source. The electrical bias was applied using a Keithley 4200 semiconductor analyzer. The laser light was focused on the sample surface *via* a 50x objective lens. The spectral response of the entire system was determined with a calibrated halogen-tungsten light source. The PL signal was collected by a grating spectrometer, thereby recording the PL spectrum through the CCD (Princeton Instruments, PIXIS). All the PL spectra were corrected for the instrument response. For temperature-dependent measurements, TMD/LN chips were put into a Linkam THMS 600 chamber and the temperature was set to a constant (range from $301K$ to $83K$) during the PL measurements using a low-temperature controller with liquid nitrogen coolant.

**Acknowledgments**

We wish to acknowledge support from the ACT node of the Australian National Fabrication Facility (ANFF). Furthermore, the authors acknowledge the facilities, and the scientific and technical assistance, of the Micro Nano Research Facility (MNRF) and the Australian Microscopy & Microanalysis Research Facility (RMMF) at RMIT University. We also thank Professor Chennupati Jagadish and Professor Barry Luther-Davies from the Australian National University (ANU) for facility support. We acknowledge financial support from the China postdoctoral scholarship (grant number

2016M590806), National Natural Science Foundation of China (NSFC) (grant number 61435010, 61875138, and 61775147); The Science and Technology Innovation Commission of Shenzhen (grant number JCYJ20170818093453105); Australian Research Council (ARC) No. DE140100805 and DP180103238.

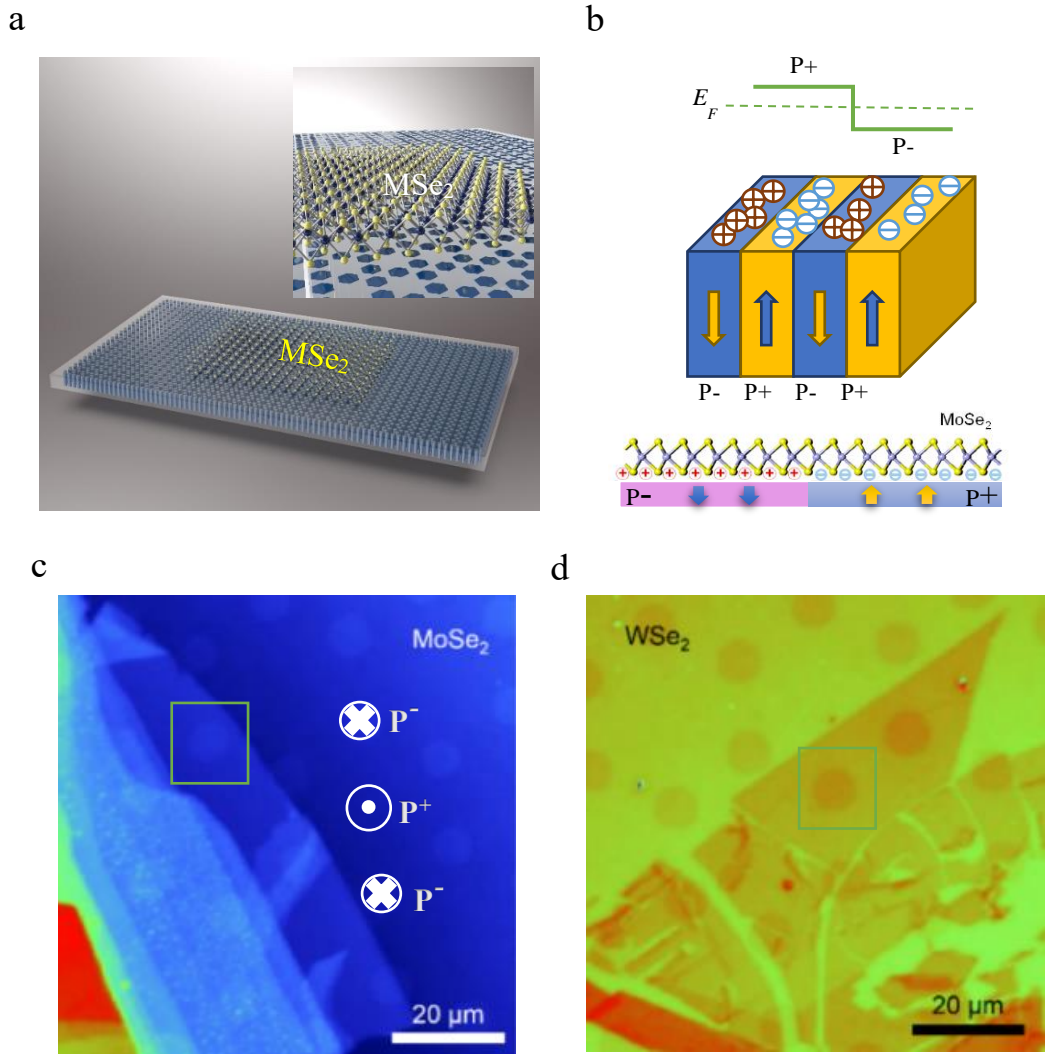

**Figure 1 | Schematic and optical images of monolayer MoSe₂ and WSe₂ on PPLN.**

**a)** Schematic of MSe₂ (M: Mo or W) on 2D domain patterned LN for spatial carrier density modulation. Both polarization states (down state P⁻ and up state P⁺) on a ferroelectric LN can directly affect the carrier density of the monolayer TMDs. **b)** Schematic diagram of the surface doping on domain patterned LN surface which can reconstruct the band of the TMD samples. For the monolayer MoSe₂ on 2D domain patterned LN, the P⁻ surface is rich of positive interface charges, which increases the electron transfer out of the MoSe₂ sample and decreases the intrinsic n-doped character of the MoSe$_2$. **c, d**) Phase-shifting interferometry (PSI) image of the exfoliated monolayer MoSe$_2$ and WSe$_2$ on 2D domain patterned LN. Domain direction was marked in the PSI image (**c**) as cross ($\otimes$) to down domain (P$^-$) and dot ($\odot$) to up domain (P$^+$). PSI images show exfoliated monolayer MoSe$_2$ with optical thickness 4 nm and 3.7 nm. The box indicated by the green line in (**c**) indicates the further PL mapping area.

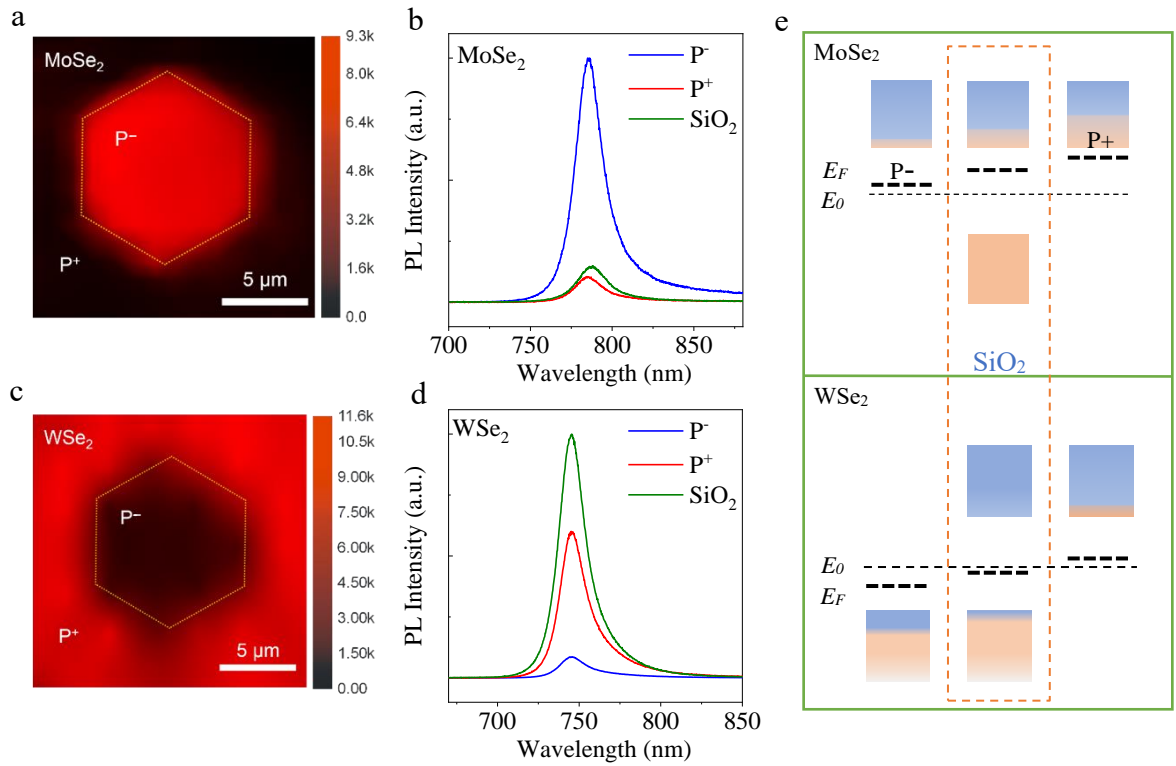

**Figure 2 | Strong photoluminescence (PL) modulation in monolayer MoSe$_2$ and WSe$_2$. a)** PL mapping of exfoliated monolayer MoSe$_2$ on a single polarized domain. The gold dash line indicates one single dipole. **b)** PL spectrums come from monolayer MoSe$_2$ on different polarized areas and SiO$_2$/Si substrate as control. **c)** PL mapping of exfoliated monolayer WSe$_2$ on a single polarized domain. **d)** PL mapping of exfoliated monolayer WSe$_2$ on a single polarized domain. **e)** Schematic plot of ferroelectricity-induced doping level modulation in TMD materials.

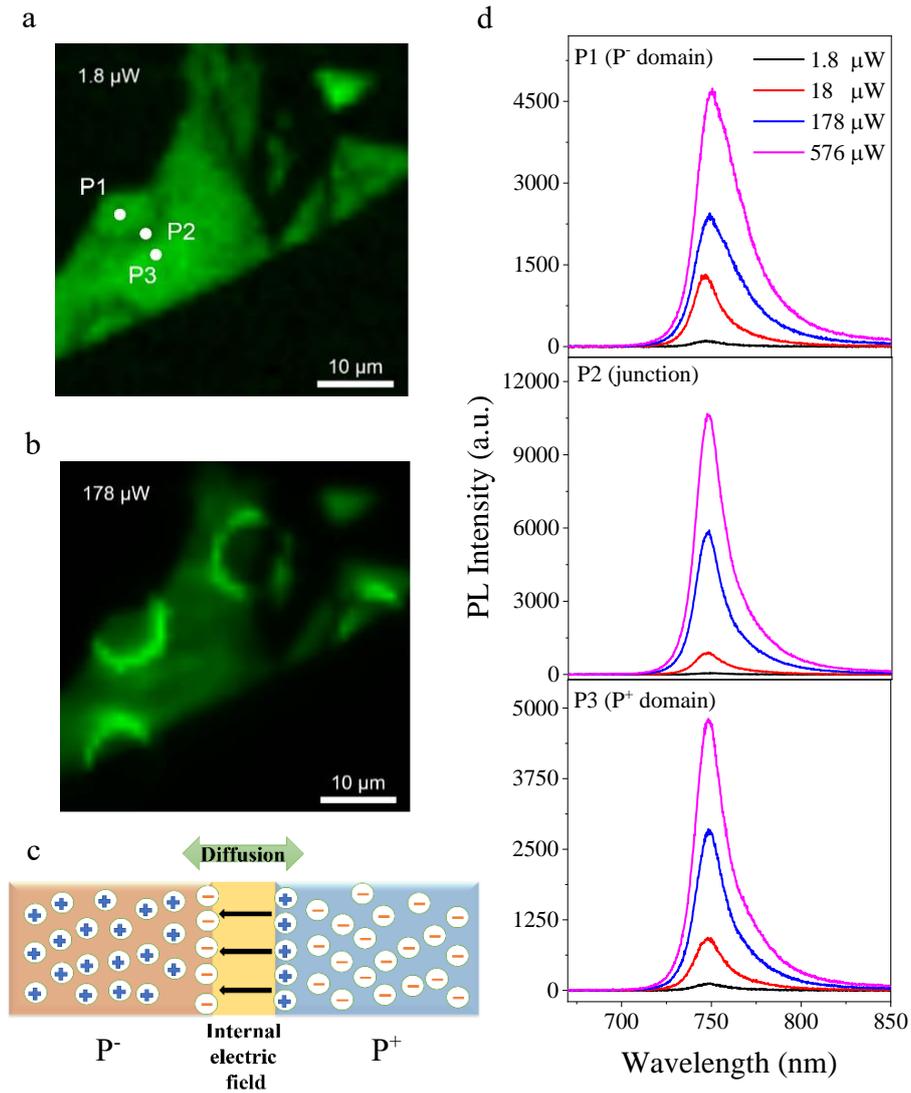

**Figure 3 | RT laser excitation power dependent PL modulation in monolayer WSe₂.**

**a, b)** measured spatial PL mapping with various excitation power of 1.8 μW (b). **c)** scheme plot of exciton diffusion on junction area by internal electric field. **d)** extracted PL spectra from PL mapping with various spatial position as P⁻ (P1), junction area (P2) and P⁺ (P3) at different excitation laser power.

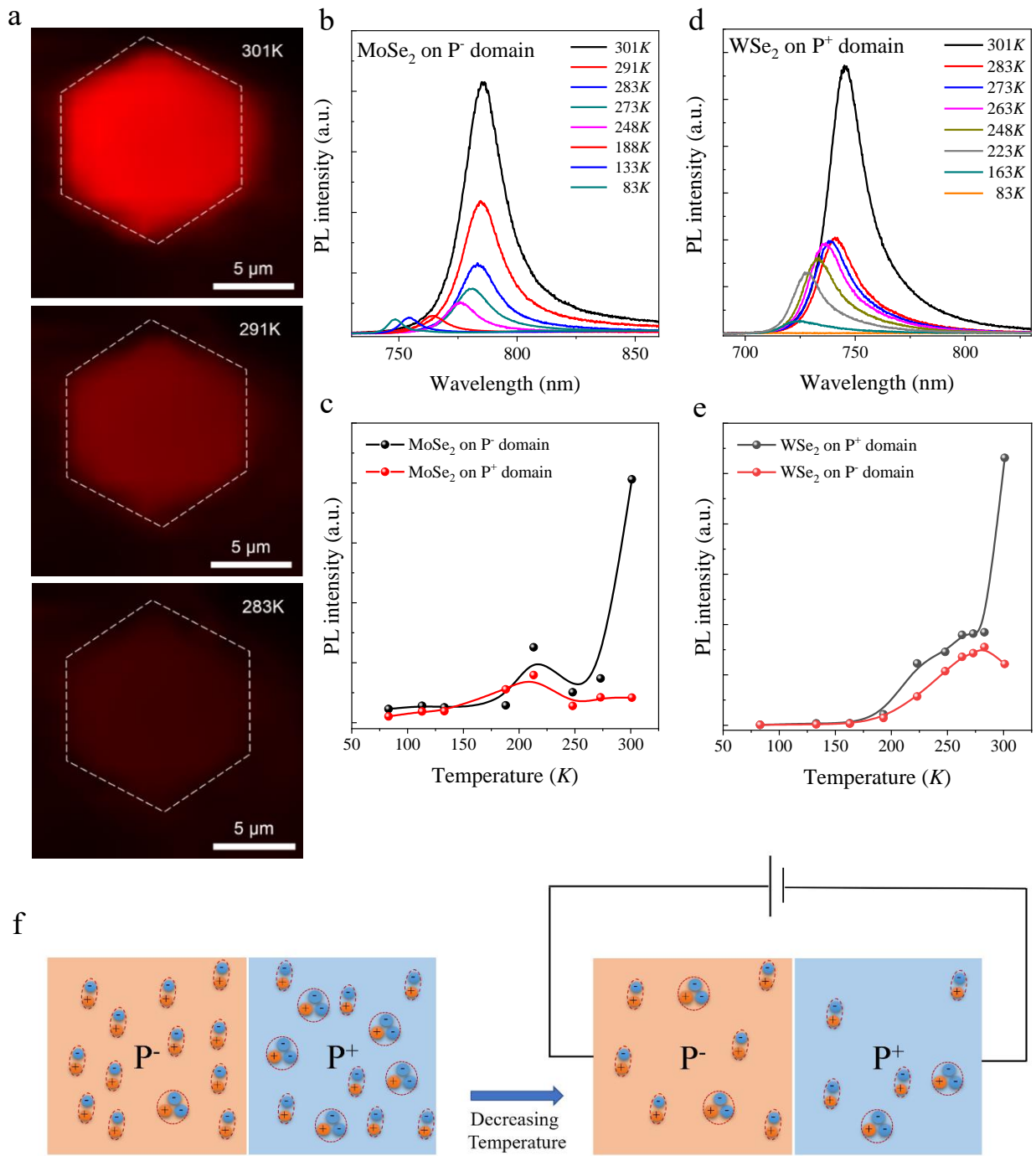

**Figure 4 | Temperature dependent PL modulation in monolayer MoSe₂. a)** PL mapping of exfoliated monolayer MoSe$_2$ on a single polarized domain under 301 *K*, 291 *K* and 283 *K*. The white dash lines in (a) indicate the single hexagonal dipole in the domain engineered LN substrate. **b)** PL spectrums of monolayer MoSe$_2$ on the P⁻

domain from 301 $K$ to 83 $K$. **c)** Integrated PL intensity of the MoSe$_2$ on P$^-$ and P$^+$ domains with decreasing temperature. **d)** PL spectrums of monolayer WSe$_2$ on the P$^+$ domain from 301 $K$ to 83 $K$. **c)** Integrated PL intensity of the WSe$_2$ on P$^-$ and P$^+$ domains with decreasing temperature. **f)** Schematic plot of the TMD samples on the domain engineered LN when decreasing the temperature.